\documentclass[aps,pra,superscriptaddress,preprint,showpacs]{revtex4}

\usepackage{graphicx}
\usepackage{graphics}
\usepackage{amssymb}
\usepackage{amsmath}
\usepackage{epsfig}
\usepackage{latexsym}
\usepackage{color}
\usepackage{rotating}
\usepackage{subfigure}

\begin{document}

\title{Single-photon opto-mechanics in the strong coupling regime}

\author{U. Akram}
\affiliation{Department of Physics, School of Mathematics and Physics, The University of Queensland, St Lucia, QLD 4072, Australia}
\author{N. Kiesel}
\affiliation{Faculty of Physics, Quantum Optics, Quantum Nanophysics and Quantum Information, University of Vienna, Austria}
\author{M. Aspelmeyer}
\affiliation{Faculty of Physics, Quantum Optics, Quantum Nanophysics and Quantum Information, University of Vienna, Austria}
\author{G. J. Milburn}
\affiliation{Department of Physics, School of Mathematics and Physics, The University of Queensland, St Lucia, QLD 4072, Australia}
\begin{abstract}
We give a theoretical description of a coherently driven opto-mechanical system with a single added photon. The photon source is modeled as a cavity which initially contains one photon and which is irreversibly coupled to the opto-mechanical system. We show that the probability for the additional photon to be emitted by the opto-mechanical cavity will exhibit oscillations under a Lorentzian envelope,  when the driven interaction with the mechanical resonator is strong enough. Our scheme provides a feasible route towards quantum state transfer between optical photons and micromechanical resonators.
\end{abstract}

\maketitle
\section{Introduction}
A significant engineering discipline has been built around the ability to fabricate micron- and nano-scale opto-mechanical systems of extraordinary variety. The emerging field of quantum optomechanics extends this ability towards a fully quantum domain enabling a new scientific discipline that aims to establish mechanical resonators as novel systems for quantum science. In combination with quantum optics techniques and new fabrication methods, highly nonclassical states of motion, such as a vibrational energy eigenstate, squeezed states and even entangled states can be prepared and coherently manipulated~\cite{optomechanics-review-Science,focus-issue,APS-Physics}. This now provides a new approach for controlling the mutual interaction between light and mesoscopic structures, which is one of the eminent goals in quantum information science~\cite{QIS_roadmap} and of importance for fundamental experiments at the quantum-classical boundary~\cite{Schroedinger}.

Almost all previous investigations in opto-mechanics have presupposed conventional optical sources, well described by statistical mixtures of coherent states.  Some early theoretical work considered the possibility of using squeezed light in an opto-mechanical setting\cite{caves} and advanced LIGO may make some of these suggestions an engineering reality\cite{LIGO}. More generally speaking, most of the current proposals to achieve (opto-)mechanical quantum states are restricted to the class of Gaussian states. To go beyond this regime requires additional non-linearities, either in the interaction or in the measurement process. One example is the use of single photons to prepare macroscopic mechanical superpositions~\cite{Marshall2003,Pikovsky2008}. Current opto-mechanical systems, however, still exhibit couplings below the necessary single-photon interaction strength. In this paper we propose a scheme that allows to achieve single-photon optomechanics in presently available systems. The main idea is to enhance the single-photon coupling strength by the presence of a strong pump field. It has recently been shown both in theory~\cite{Marquardt2007,Dobrindt2008} and experiment~\cite{Groblacher} that this allows to enter the strong coupling regime of an opto-mechanical system. We show that in such a case, even for small intrinsic single-photon coupling, a single photon excitation of the cavity can be reversibly transferred to the vibrational motion of a mechanical resonator. We study the dynamics of this process and show that it can be detected as temporal oscillations in the cavity emission. This is in close analogy to optical three-wave mixing, where the pump field converts excitations in the optical signal mode (here: the cavity photons) into excitations in the optical idler mode (here: the vibrational phonons) and vice versa. 

We consider a single-mode optical cavity of length $L$, frequency $\omega_c$ and linewidth $\kappa$, with a moving end mirror that is modeled as a simple mechanical resonator with mass $m$ and resonant frequency $\omega_m$. We also assume operation in the resolved-sideband regime, i.e. $\kappa< \omega_m$.  The cavity is strongly driven with a coherent pump field at frequency $\omega_L$.  We describe the interaction through a linearized treatment that is expanded around the steady state field amplitude in the cavity, which would arise in the absence of the opto-mechanical interaction. In addition to the coherent driving field, the cavity is also driven by a single-photon source. This is intended to be a sequence of pulses with one and only one photon per pulse, however it will suffice to consider only a single pulse for the purposes of the calculation presented here. 

\begin{figure}[htbp] %  figure placement: here, top, bottom, or page
   \centering
   \includegraphics[scale=0.7]{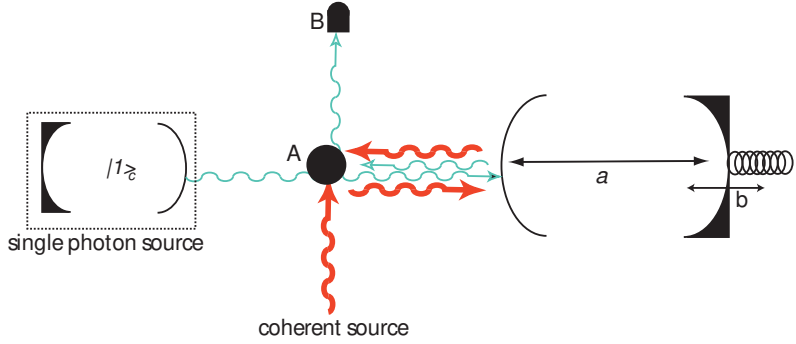} 
   \caption{Scheme of the setup. A single-sided opto-mechanical cavity of frequency $\omega_c$ (right; depicted as Fabry-Perot cavity with a moving end mirror) is driven by both a single-photon pulse and by a continuous-wave (CW) coherent pump source. The single photons are emitted from a cavity (left) at resonance with the opto-mechanical cavity, while the CW pump source is detuned from $\omega_c$. The pump beam produces a strong coherent field inside the cavity that enhances the coupling between the added single photons and the mechanical resonator analogous to optical three-wave mixing. An optical circulator (A) allows one to separate the photons emitted from the steady state coherent field inside the cavity from the added photons produced by the opto-mechanical interaction. The circulator may be realized by a coherent state displacer or by using an additional optical cavity as spectral separator, in which the driving field is transmitted while the added single photons are reflected. The latter are then detected on a photodetector (B). which are then detected on a photodetector (B)}
   \label{fig1}
\end{figure}

To model the single-photon source we include a source cavity of frequency $\omega_s$ and of decay rate $\gamma$, which at $t=0$ is prepared in a single photon state (Figure~\ref{fig1}). The coupling between the source cavity and the opto-mechanical cavity is irreversible and can be described using the cascaded systems approach \cite{Car,Gar}.  In this way we obtain a source that produces a single-photon pulse with a Lorentzian line shape. In the following, the single-photon source cavity is on resonance with the opto-mechanical cavity.  The additional coherent laser driving field of frequency $\omega_L$ enhances the opto-mechanical radiation pressure coupling. This field is spectrally detuned from the cavity resonance by multiples of the mechanical resonance frequency so that motional side bands can be addressed. Our objective is to demonstrate coherent exchange of the added single-photon optical excitation with the vibrational excitation of the mechanical mirror. To this end we compute the dynamics of the mean added photon number in the cavity, $n_a$, which then determines the single-photon detection rate at detector $A$ as $\kappa n_a$ where the cavity damping rate is $\kappa$. We will find that for strong coupling this detection probability oscillates due to the coherent exchange of the single photon with the mechanical phonon number. 

The interaction between the cavity field of the source and the mechanical  motion of the end mirror is via the standard radiation pressure coupling\cite{Law1995,Jacobs}
\begin{equation}
H_{rp}=\hbar G a^\dagger a(b+b^\dagger)
\end{equation}
with the coupling rate 
\begin{equation}
G=\frac{\omega_c}{L}\sqrt{\frac{\hbar}{m\omega_m}}
\end{equation}
where $m$ is the effective mass of the moving mirror. For current opto-mechanical systems $G$ is small compared to the cavity linewidth $\kappa$. However, a larger effective interaction is obtained by driving the cavity with a strong coherent laser field. In this case the amplitudes of both the cavity field and the mechanical motion will be displaced by their steady state amplitudes $\alpha_0$ and $\beta_0$, respectively. One can then linearize the radiation pressure force in the shifted reference frame to get~\cite{Groblacher} 
\begin{equation}
H=  \hbar\Delta a^\dagger a+\hbar\omega_m b^\dagger b+\hbar g(a+a^\dagger)(b+b^\dagger)
\label{strong-coupling}
\end{equation}
where $g=G\alpha_0$ (we have fixed the phase for the coherent drive to make this a real variable) is the effective coupling strength, $\Delta=\omega_c-\omega_L+2g$ is the effective cavity detuning, and $a,a^\dagger$ and $b,b^\dagger$ are the annihilation and creation operators for the displaced cavity field and for the motion of the mirror mechanical resonator, respectively. In the absence of damping the Heisenberg equations of motion are linear and may be solved by the method of normal modes. The normal mode frequencies are given by
\begin{equation}
\omega_{\pm}^2=\frac{1}{2}\left (\Delta^2+\omega_m^2\pm\sqrt{(\Delta^2-\omega_m^2)^2+16g^2\Delta\omega_m}\right )
\end{equation}
The normal mode splitting occurs for any $g >0$ in the undamped case, but when damping is included a minimum coupling strength is required and given by $g>\kappa$ (see the Supplementary Information in \cite{Groblacher}). In the undamped case, oscillatory solutions will occur in the sought single-photon detection probability when $\omega_{\pm}$ are real, i.e.  $4g^2\leq \Delta\omega_m$. 

 In the limit that $4g^2 << \omega_m\Delta$ we can chose the detuning $\Delta$ to make particular resonant terms ($\Delta=\pm\omega_m$) dominate the interaction.  To identify the resonant terms we first move to an interaction picture defined by the free Hamiltonian, $H_0= \hbar\Delta a^\dagger a+\hbar\omega_m b^\dagger b$. The resulting time-dependent Hamiltonian may be approximated by a time-independent Hamiltonian using the appropriate resonance condition. For example, if $\Delta=\omega_m$, the interaction Hamiltonian can be approximated by the {\em red sideband} coupling
\begin{equation}
H_r=\hbar g(ab^\dagger+a^\dagger b)
\label{red-sb}
\end{equation}
which leads to cooling of the mechanical motion if the optical cavity is rapidly damped\cite{Marquardt,Wilson-Rae,Genes}. The validity of the rotating wave approximation implicit in this `side-band' Hamiltonian depends on the ratio $2g/\Delta << 1$. We note that in general the coupling constant can be quite large\cite{Groblacher}, and one may have to take the full interaction, Eq.(\ref{strong-coupling}) into account. This can lead to entanglement between the optical and mechanical degrees of freedom\cite{Vitali}, and also to heating of the mechanical motion\cite{Dobrindt2008} and eventually to an instability of the steady state resulting in self sustained oscillation on a limit cycle\cite{Ludwig}. 

The red side-band approximation, Eq.(\ref{red-sb}) describes a reversible swap of a single-photon excitation from the cavity field to the mechanical system. In the general case this exchange will not be perfect due to the counter rotating terms $ab$ and $a^\dagger,b^\dagger$. Before the single photon enters the cavity, the opto-mechanical system is in a dynamical steady state with a large circulating power in the cavity due to the coherent field $\alpha_0$. On top of this, there will be additional photons due to the excitation of both the cavity and the mechanical resonator mainly by the counter rotating terms. A single photon then enters the cavity at a random time and the system moves away from the steady state through damped oscillations (provided $4g^2\leq \Delta\omega_m$) corresponding to exchange of energy between the cavity and the mechanical resonator. This additional excitation can be lost either through the mechanical damping or through the end mirror of the cavity. Note that in the latter case the emitted photon can in principle be detected via photon counting, provided it can be distinguished from the coherent component exiting the cavity (see Section \ref{discussion}). If the coupling is large enough the excitation can be exchanged a number of times between the cavity and the mechanics before being lost. Such an oscillation will modulate the detection rate for the photons leaving the cavity and will hence leave an unambiguous signature of the coherent exchange of energy between the cavity and the mechanical resonator. Eventually this detection rate will return to zero as the opto-mechanical system returns to the steady state.

\section{Master equation for cascaded systems.}
The interaction picture master equation describing the interaction between the system is 
\begin{eqnarray}
\frac{d\rho}{dt} & = &-i\Delta[a^\dagger a,\rho]-i\omega_m[b^\dagger b,\rho] -ig[(a+a^\dagger)(b+b^\dagger),\rho]+\kappa{\cal D}[a]\rho+\gamma{\cal D}[c]\rho\\\nonumber
& & +\mu(\bar{n}+1){\cal D}[b]\rho+\mu\bar{n}{\cal D}[b^\dagger]\rho-i\Delta[c^\dagger c,\rho]+\sqrt{\gamma\kappa}\left ([c\rho,a^\dagger]+[a,\rho c^\dagger]\right )
\end{eqnarray}
where the interaction picture is defined by the coherent driving laser. Here, $\mu$ is decay rate of the mechanical system resonator, $\bar{n}$ is the mean thermal excitation of the mechanical environment at frequency $\omega_m$ and  $c,c^\dagger$ are the annihilation and creation operators for the field of the source cavity.

To demonstrate a successful state transfer we calculate the value of $\langle a^\dagger a\rangle$ as the count rate for the photon emitted from the cavity  is proportional to this quantity. As the Hamiltonian is at most quadratic in the field amplitude operators,  a closed system of equations can be obtained for the second order moments. To this end, we define the correlation matrix
 \begin{equation}
 C(t)=\langle \vec{A}(t)\vec{A}^T(t)\rangle
 \end{equation}
 where $\vec{A}^T =(a(t), a^\dagger(t), b(t), b^\dagger(t))$. This obeys the following system of equations, 
 \begin{equation}
 \frac{dC(t)}{dt}=KC(t)+C(t)^TK^T-\sqrt{\gamma\kappa} N(t)
 \label{moments}
 \end{equation}
 where 
 \begin{equation}
 K=\left (\begin{array}{cccc}
 		-\tilde{\kappa} & 0 & -ig & -ig\\
		0 & -\tilde{\kappa}^* & ig & ig\\
		-ig & -ig & -\tilde{\mu} & 0\\
		ig & ig & 0 & -\tilde{\mu}^*\end{array} \right ) 
		\end{equation}
with $\tilde{\kappa}=i\Delta+\kappa/2$ and $\tilde{\mu}=i\omega_m+\mu/2$, and the {\em noise matrix} is defined by 
\begin{equation}
N(t)=\left (\begin{array}{cccc}
		\langle ac\rangle & \langle a^\dagger c+ac^\dagger\rangle & \langle cb\rangle & \langle cb^\dagger\rangle\\
		 \langle a^\dagger c+ac^\dagger\rangle & \langle a^\dagger c^\dagger\rangle & \langle bc^\dagger\rangle & \langle c^\dagger b^\dagger\rangle\\
		 \langle cb\rangle &\langle bc^\dagger\rangle & 0 & 0\\
		 \langle cb^\dagger\rangle & \langle c^\dagger b^\dagger\rangle & 0 & 0\end{array}\right )
		 \end{equation}
The noise matrix elements obey a separate set of equations given by
\begin{equation}
\frac{d}{dt}\left (\begin{array}{c}
			\langle ac\rangle \\
			\langle a^\dagger c\rangle\\
			\langle bc\rangle\\
			\langle b^\dagger c\rangle\end{array}\right ) =\left (\begin{array}{cccc}
			-\sigma & 0 & -ig & -ig\\
			0 & -(\kappa+\gamma)/2 & ig & ig\\
			-ig & -ig & -\tau_+ & 0\\
			ig & ig & 0 & -\tau_-^* \end{array}\right )\left (\begin{array}{c}
			\langle ac\rangle \\
			\langle a^\dagger c\rangle\\
			\langle bc\rangle\\
			\langle b^\dagger c\rangle\end{array}\right )-\sqrt{\gamma\kappa}\left (\begin{array}{c}
			\langle c^2\rangle \\
			\langle c^\dagger c\rangle\\
			0\\
			0\end{array}\right )
			\end{equation}
			 (and the complex conjugate equations). We have defined $\sigma=2i\Delta+(\kappa+\gamma)/2$ and $\tau_\pm=i(\omega_m\pm\Delta)+(\mu+\gamma)/2$.  The equations of motion for the source cavity alone are
\begin{eqnarray}
\frac{d\langle c^2\rangle}{dt} & = & -(2i\Delta+\gamma)\langle c^2\rangle\\
\frac{d\langle c^\dagger c\rangle}{dt} & = & -\gamma\langle c^\dagger c\rangle	 
\end{eqnarray}
If there is a number state, $|n\rangle_c$, prepared in the source cavity at $t=0$, one immediately sees that $\langle c^2\rangle(t)$ is zero for all time, while $\langle c^\dagger c\rangle(t)=ne^{-\gamma t}$. Note that the case for an initial coherent state, $|\beta\rangle$, in the source cavity is different, as for that case $\langle c^2\rangle(t)=\beta^2 e^{-(2i\Delta+\gamma)t}, \langle c^\dagger c\rangle(t)=|\beta|^2e^{-\gamma t}$, which makes the dynamics dependent on the phase of $\beta$. The number state case, in contrast, has no similar phase reference. The equation for the noise matrix may be solved directly in both cases and substituted into the equation for the correlation matrix $C$. Note that the intracavity photon number will depend on the time dependent correlations between $a$ and $c$. This starts at zero, rises to a maximum as the single photon excitation begins to grow in the cavity and then decays to zero. 

 We discuss our results in direct comparison with present experimental parameters. An experiment reported by Gr\"{o}blacher et al\cite{Groblacher},  demonstrated the strong coupling regime with a linewidth of the optical cavity and the mechanical resonator of $\kappa=2\pi\times 215$kHz and $\mu=2\pi\times 140$Hz, respectively and with an effective coupling strength of $g=2\pi\times 325$kHz.  The mechanical resonator frequency was $2\pi\times 947$kHz. In units such that $\kappa=1$ these are equivalent to $\omega_m=4.4,\mu=6.5\ \times 10^{-4}, g=1.5$. In the following we set the detuning $\Delta=1.02\omega_m$ and we choose the source cavity to be nearly mode matched to the opto-mechanical cavity $\gamma=0.9$. Before the single photon source is turned on, we assume that the opto-mechanical system has reached a steady state, which then become the initial conditions when the source is turned on; i.e. we need to find the steady state solutions of Eq(9) with $\gamma=0$. The steady state solution for the opto-mechanical covariance matrix is given by $KC_\infty+C_\infty^TK^T=0$. 
 
A great deal of experimental effort is going into schemes to enable the mechanical motion of the mirror to be cooled to its ground state, and we first consider this case. Practically, this will occur due to optical cooling via the red-detuned pump beam if we assume that the rate of heating due to the interaction of the mechanical resonator with its environment is small . We therefore set $\bar{n}=0$ and $\mu=0.001$ in the equations of motion. Furthermore, we choose the single-photon linewidth $\gamma = 0.9$.   

\begin{figure}[h]
   \centering
   \includegraphics[scale=0.6]{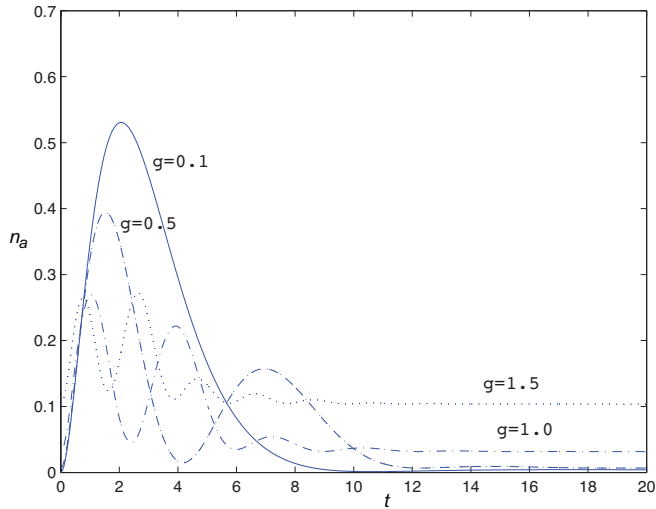}
   \caption{The mean photon number in the opto-mechanical cavity versus time: $\bar{n}=0, \gamma = 0.9, \kappa = 1.0,\mu=0.001,\ \omega_m=4.4, \Delta= 1.02\omega_m, \ \omega_m=4.4$ and four values of $g$.}
   \label{fig2a}
\end{figure}

\begin{figure}[h]
   \centering
   \includegraphics[scale=0.6]{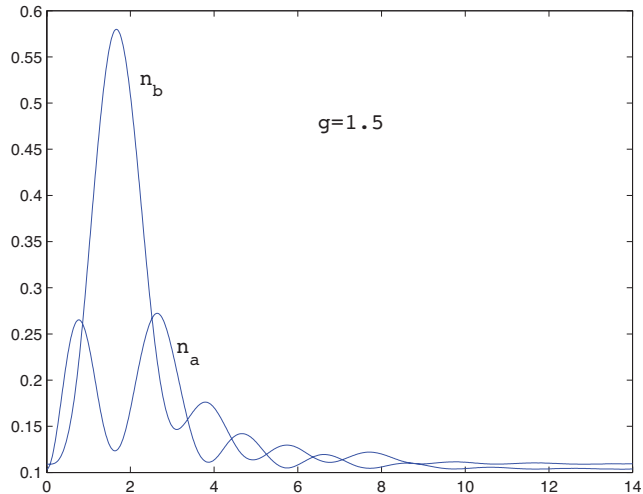}
   \caption{(a) The mean photon number in the opto-mechanical cavity and the mean phonon number in the mechanical resonator versus time for $g=1.5$, $\bar{n}=0, \gamma = 0.9, \kappa = 1.0,\mu=0.001,\ \omega_m=4.4, \Delta= 1.02\omega_m, \ \omega_m=4.4$.}
   \label{fig2b}
\end{figure}

\begin{figure}[h]
   \centering
   \includegraphics[scale=0.6]{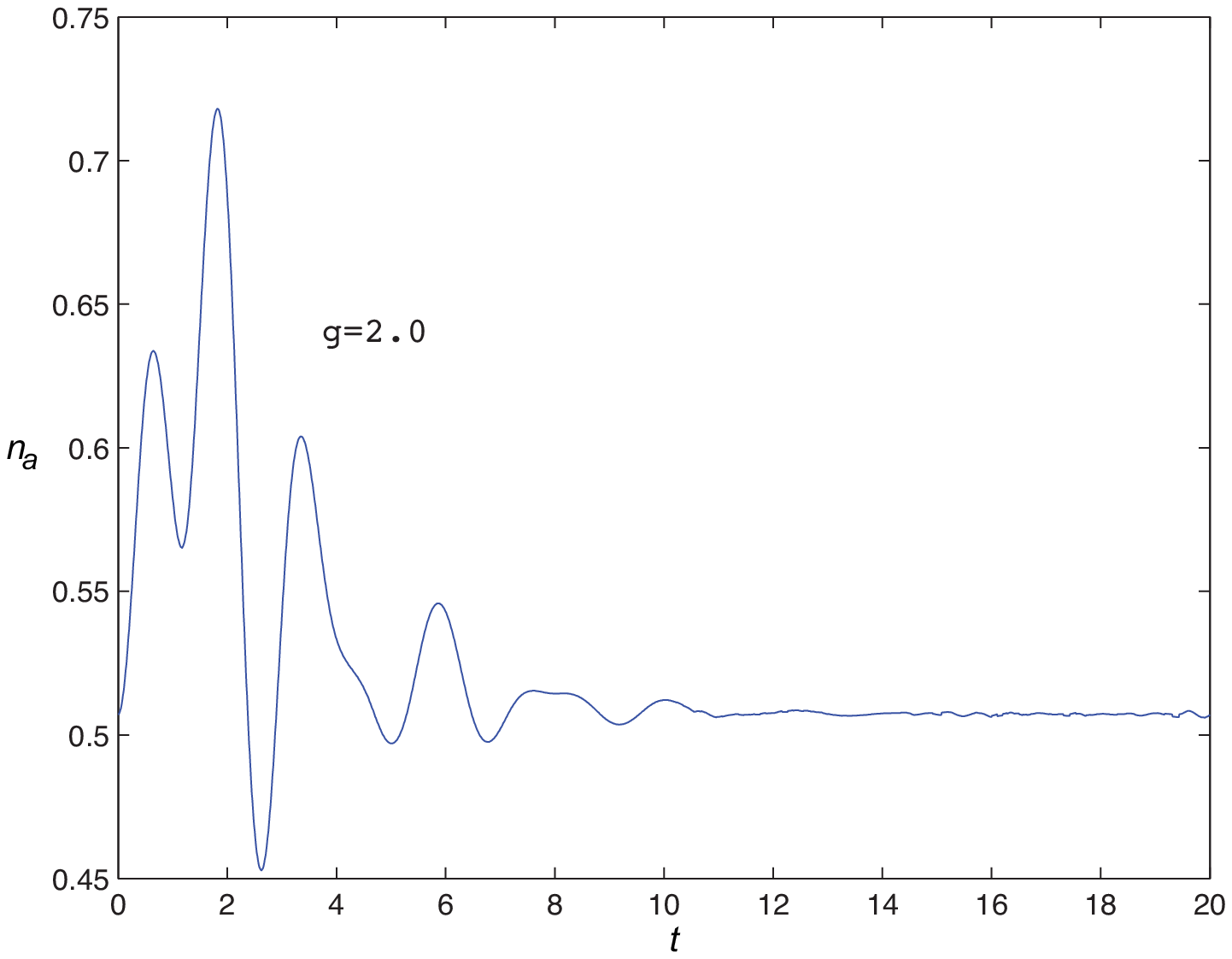}
   \caption{The mean photon number in the opto-mechanical cavity versus time: $\bar{n}=0, \gamma = 0.9, \kappa = 1.0,\mu=0.001,\ \omega_m=4.4, \Delta= 1.02\omega_m, \ \omega_m=4.4$ with $g=2.0$.
}
   \label{fig3}
\end{figure}

Figure \ref{fig2a} shows the time-dependence of the mean cavity photon number for various coupling strengths $g<\kappa,\Delta$. For $g << \Delta$ we simply recover the statistics of the cavity decay as no significant opto-mechanical coupling takes place. Increasing $g$ such that $\kappa < g < \Delta$ we observe revivals in the detection probability, which arise because the single-photon excitation is exchanged coherently between the opto-mechanical cavity and the mechanical resonator. This can be seen directly in Figure~\ref{fig2b} where we plot the simultaneous evolution of both the intra cavity mean photon number and the mean phonon number in the mechanical resonator for the case of $g=1.5$. It might be noted that the photon number and phonon number oscillations are not  $\pi$ out of phase in the first phase of the evolution as one might expect if the cavity was started with exactly one photon at $t=0$. For short times, this is due to the dynamics of the single photon source excitation of the cavity, on top of the photon-phonon interactions: the dynamics in Eq.\ref{moments} depends explicitly on the correlations between the source and the optomechanical cavity, $\langle ac^\dagger+a^\dagger c\rangle$. At later times, the mean cavity photon number does not peak at the same time as the minimum in mean mechanical phonon number because the decay rate of the cavity is very much greater than the mechanical decay rate.  For $g > \kappa, \Delta$, see Figure \ref{fig3}, the oscillations persist but additional frequencies appear due to normal mode splitting and there is an excitation of the dressed opto-mechanical system, analogous to heating. 


\begin{figure}[h!]
   \centering
   \includegraphics[scale=0.6]{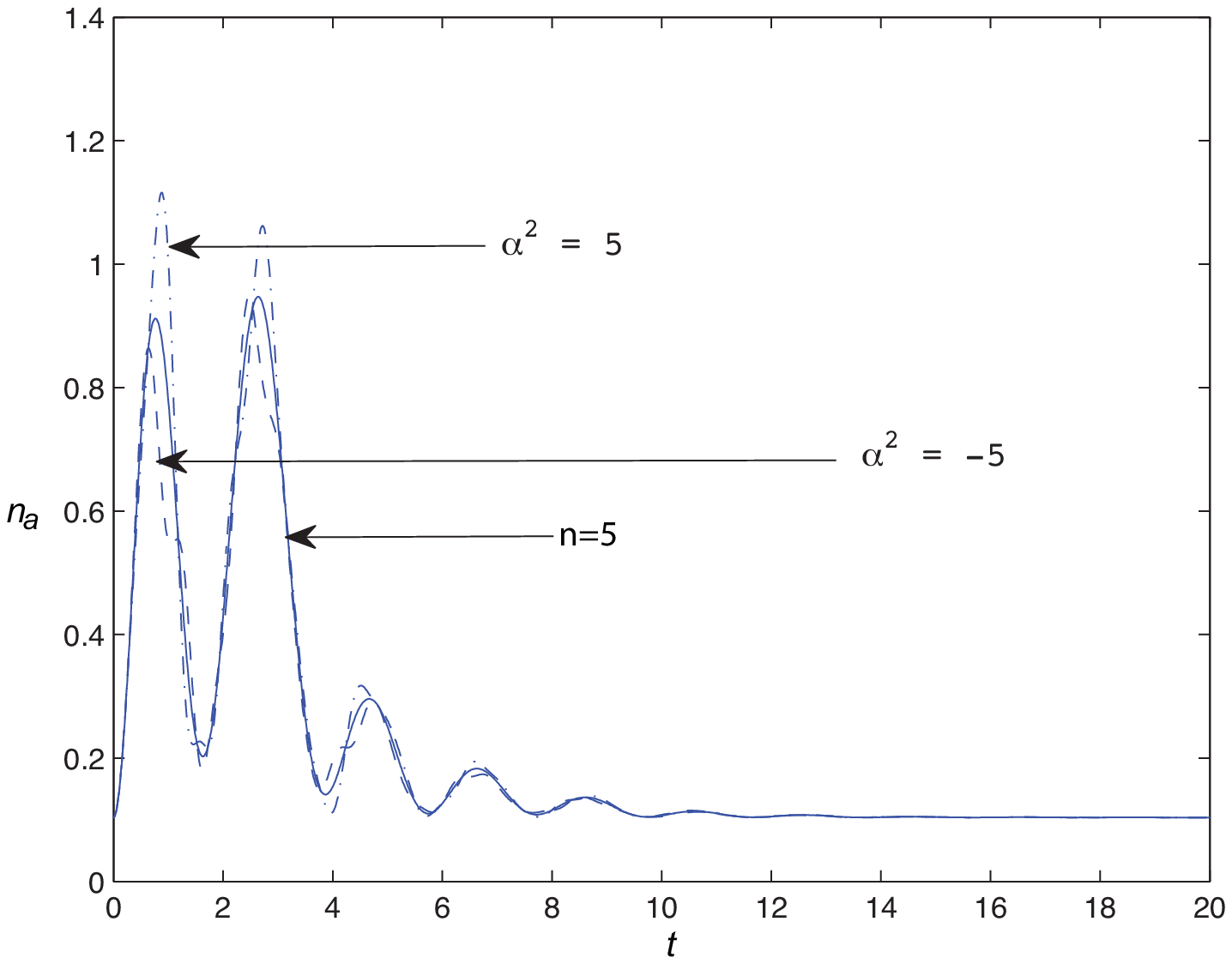} 
   \caption{The mean photon number in the opto-mechanical cavity versus time, contrasting the case of an initial Fock state source and an initial coherent state source. In all cases $g=1.5$. The mechanical resonator is at zero temperature, $\bar{n}=0$, and the opto-mechanical cavity starts with no photons. The source cavity is prepared in a Fock state $n=5$ (solid line) or a coherent state with amplitude $\sqrt{5}$ ( dashed-dot) and $\sqrt{5}i$ (dashed) }
   \label{fig4}
\end{figure}

As discussed above we expect the dynamics for an initial Fock state in the source cavity to differ from that for an initial coherent state with the same mean photon number. This is shown in figure \ref{fig4} where we compare the dynamics for a Fock state in the source cavity $n=5$, and two coherent states, $|\alpha\rangle$ with $\alpha=\sqrt{5}$ and $\sqrt{5}i$. 

\begin{figure}[h]
   \centering
 \includegraphics[scale=0.6]{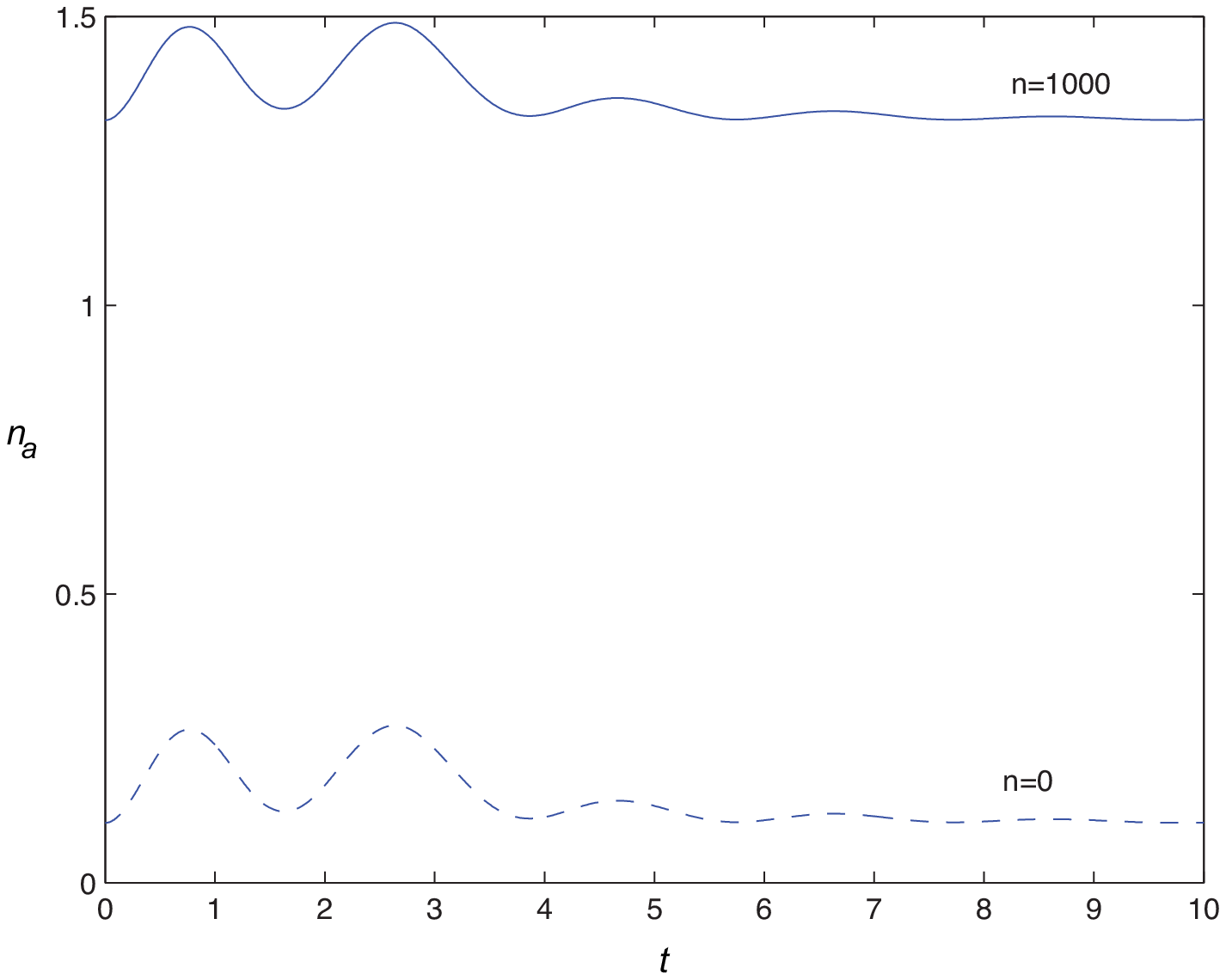} 
   \caption{The mean photon number in the opto-mechanical cavity versus time showing the effect of the thermal driving of the nanomechanical resonator for $\bar{n}=1000$,    $\gamma = 0.9, \kappa = 1.0,\mu=0.001$ and $g=1.5$. The case of $\bar{n}=0$ is shown for comparison in the dashed line. 
}
   \label{fig5}
\end{figure}

We finally include coupling of the mechanical resonator to a nonzero temperature heat bath, here $\bar{n}=1000$ (see Figure~\ref{fig5}). This can be accomplished by starting from different initial conditions that take into account that, prior to the single-photon injection, the optical cavity and the mechanical resonator are in thermal equilibrium.  While the signal is of the same magnitude as in the zero-temperature case, there is an added noise that corresponds to the steady-state thermal occupation of the mechanical oscillator.

\section{Discussion and Conclusion}
\label{discussion}
The calculation we have presented is based on a linearization of the intensity dependent force acting on the mechanical element, around a strong coherent steady state field inside the cavity. This means that the average photon numbers we have calculated are {\em in addition to} a coherent steady state field. In order to detect the added photon number due to the linearized interaction, over and above the steady state coherent field inside the cavity, we need to subtract off the known steady state field amplitude $\alpha_0$ by displacing the output field amplitude from the cavity before  sending it to a photodetector. Such displacements can, for example, be  done by mixing the output field from the cavity with a local oscillator coherent field (split off from the driving laser) on a beam splitter with very high reflectivity, see for example \cite{Furasawa}.  Another possibility is to spectrally separate the two components by an additional filtering cavity at resonance with the driving field. With the coherent amplitude displaced away, the photon detection rate is proportional to $\kappa$ times the mean photon number as presented in figures 2-6. 

We have modeled the single photon source as a single cavity initialized with one photon. In order to sample the mean photon number in the cavity, the single photon source cavity needs to be re-prepared. In reality, a single photon source is either a pulsed or a heralded source with one and only one photon per trigger event\cite{milburn-SPS}. The model we have used can apply to these case provided that the period between pulses is sufficiently long that the opto-mechanical system can return to steady state after detection of the single photon emitted from the cavity between each pulse. In addition, our cavity model assumes an exponential temporal pulse shape. These assumptions are however consistent with new narrow-linewidth single-photon sources that have been developed in the context of atom-light interfaces\cite{Melholt}. Yet, the actual pulse shape is not very important provided it is matched reasonably well to the opto-mechanical cavity line-width. Finally, the timing information in heralding the single photon further helps to reduce noise in the experiment by a gated detection scheme.

We have proposed a novel scheme that allows the coherent exchange of single-photon excitations of an optical cavity with a micromechanical resonator. The single-photon coupling is enhanced by a strong pump field that mediates the state transfer, in close analogy to optical three-wave mixing. A clear signature of the state transfer between light and mechanics is the oscillation of the added-photon emission probability from the cavity. The scheme can be realized with state-of-the-art optomechanical systems that operate sufficiently close to the quantum ground state.  This provides the basis for storage and interacting of optical photons in/via mechanical structures. Note that a similar idea has recently been suggested to create mechanical superposition states~\cite{Cirac}.  

\acknowledgements
We thank M. Lukin for discussions. GJM wishes to acknowledge the support of the University of Vienna and the Australian Research Council. UA acknowledges the support of the Australian Research Council and the University of Queensland.  MA acknowledges support from the Austrian Science Fund (projects P19539, START), the European Commission (project MINOS), the European Research Council (ERC QOM) and the Foundational Questions Institute FQXI. NK acknowledges support by the Alexander-von-Humboldt Foundation.

\end{document}